# A particle mechanism for the index of refraction

**Marcel URBAN**
LAL, Univ Paris-Sud, IN2P3/CNRS, Orsay, France

**Abstract.** We propose to go away from the electromagnetic wave description of light and to explain through a purely corpuscular and neutral approach, the phenomenon of the slowing down of light in a transparent medium. The quantum predictions and ours are compared against experimental refractive indices. In our framework the optical Kerr phenomenon finds a very natural interpretation and its numerical values are obtained easily.

## 1. Introduction

Leon Foucault showed experimentally [1] that light slows down in water with respect to vacuum. More generally visible light velocity is reduced in transparent media. In a homogeneous material this is described by a single constant: the index of refraction n, which is the ratio of the velocity of light in vacuum to its velocity in the medium.

Molecules have energy states quantized and in a gas these states are very well separated as opposed to solids where they can merge and form continuous bands. A photon propagating through a gas is absorbed by a molecule only if its energy corresponds to an allowed difference of energies in this molecule. For instance if the first excited state is 10 eV above the ground state, visible photons of a few eV should do nothing to these molecules.

Why then do we have an index of refraction for a continuum of photon energies? Or put differently, how come a photon whose energy does not fit an atomic line can be influenced by the medium?

We begin to recall the wave approach to the origin of the index of refraction. Then we propose another mechanism for the slowing down of the photons. The idea is to push the particle model of light as far as possible without using wave's properties. We describe our model of the interaction between a photon and a molecule in matter and we undertake a comparison between the two approaches.

In this corpuscular framework, we predict that there will be a dispersion of the arrival times for mono energetic photons traversing a sheet of matter. This could have been already seen in 1987.

Another benefit of the corpuscular approach is that we understand the optical Kerr phenomenon easily and that we derive its magnitude in gases and in condensed matter.

## 2. The usual explanation for the index of refraction

To the best of our knowledge there is only one explanation for the origin of the index of refraction and a clear description is made in [2] and in [3]. The first step considers an electromagnetic monochromatic plane wave falling on a sphere of dielectric whose diameter is very much smaller than the wave length $\lambda$. The electric field of the incident wave induces a vibrating electric dipole in the sphere which emits a spherical wave of the same frequency as the one of the incoming wave. This is the incoherent Rayleigh scattering on a single scattering centre. The downstream amplitude of the incident plane wave is slightly reduced but there is no change in its velocity.

The second step considers a continuous slab of dielectric having a small depth and a large extension perpendicular to the axis of propagation of the plane wave. The slab of dielectric is decomposed into small volumes, as compared to $\lambda^3$, where the incident plane wave induces electric dipoles. These elementary vibrating dipoles emit secondary spherical waves according to the Rayleigh incoherent cross section. The reason to consider a very thin slab of matter is that the sum of these secondary waves on one particular elementary dipole is negligible compared to the incident light amplitude. The driving of the elementary dipoles is then proportional only to the amplitude of the incident light. All these secondary waves add up coherently downstream to the incoming light. The coherent integration of these secondary waves is phase shifted by $\pi/2$ with respect to the incident wave. Since the thickness of the slab is small, the amplitude iA of this integration has a smaller modulus than the amplitude of the incident plane wave taken as 1. The sum, downstream, of the incident light and of the secondary waves takes the form 1+iA which is approximated as $e^{iA}$. Thus instead of being 1 the incident wave is transformed into $e^{iA}$ which displays a phase shift interpreted as a delay in time due to a smaller speed in the slab of dielectric. This at last gives the index of refraction.

The strength of the induced electric dipoles is given by their amount of polarisation under the influence of the incident electric field. This approach was envisaged for a continuous distribution of matter, and when the atomic nature of matter was established it became possible to have a microscopic model for the incoherent polarisation. At the same time we have the difficulty to go from a discrete to a continuum. The continuum is necessary in order to use the framework of the coherent integration producing the phase shift and thus explaining the index of refraction. The small dielectric sphere of the incoherent Rayleigh scattering is replaced by the molecule. The electric field of the incident wave sets the electrons in the molecule into a dipolar periodic motion and these electrons are supposed to reemit light spherically and at the same frequency as the one of the incoming wave.

Any approach to the prediction of the index of refraction is based upon the calculation of the average electric polarisation of the molecule, produced by the incident electric field of the plane wave. Then, as soon as possible, go to the macroscopic polarisation. Once in the continuum theory framework the dielectric constant $\varepsilon$ is obtained as a function of frequency and finally the index of refraction is predicted through the Maxwell formula: $n^2 = \varepsilon$. We will call this line of thought the Maxwell-quantum framework.

### 3. What is found in the literature?

The semi classical and quantum predictions of the index of refraction can be found in textbooks [4]. The semi classical formula for the index of refraction of a photon of energy $E_\gamma$ admits that the electrons in the molecule, behave like oscillators with discrete energies $W_i$

$$n^2 - 1 = 4\pi\, r_e (\hbar c)^2 \sum_i \frac{N_i}{W_i^2 - E_\gamma^2} \qquad \sum_i N_i = N_{elec} \qquad (3.1)$$

$N_{elec}$ is the number of valence electrons per unit volume and $r_e = 2.83\ 10^{-15}$ m, is the classical radius of the electron. $\hbar c = 197.3\ MeVfm$. The hydrogen atom has $W_0 = -13.6$ eV and $W_1 = -3.5$ eV.

When the polarisation of the molecule is calculated with the quantum mechanics, we get what we call the Maxwell-quantum formula

$$n^2 - 1 = 4\pi\, r_e N_{elec} (\hbar c)^2 \sum_i \frac{f_i}{E_i^2 - E_\gamma^2} \qquad \sum_i f_i = 1 \qquad (3.2)$$

If the ground state energy of the atom is $W_0$ then: $E_i = W_i - W_0$.
The hydrogen atom has $E_1 = W_1 - W_0 = 10.2$ eV.
Therefore the Maxwell-quantum formula differs mainly in the characteristic energies $E_i$ which are smaller than the $W_i$ of the semi classical formula.

In a condensed medium, Lorentz and Lorenz try to take into account the difference between the incident and the local electric field and suggest the following formula:

$$\frac{n^2 - 1}{n^2 + 2} = \frac{4\pi}{3} r_e N_{elec} (\hbar c)^2 \sum_i \frac{f_i}{E_i^2 - E_\gamma^2} \qquad (3.3)$$

So we understand that the mediums where we have a chance to predict correctly the index of refraction are the gases. However a surprise awaits us. All textbooks concerning the index of refraction theory spend a fair amount of pages and mathematics to first derive the semi classical formula (3.1) then the Maxwell-quantum formula (3.2) and yet no comparison with real data is made! They are in fact copying each other about this subject. A. Sommerfeld is an exception because he does compare, quickly, the classical formula with the experimental data on gaseous hydrogen [5] and he recognises that 'the theory is still very crude...'

Because of this copying we may as well go to one of the first textbook like H. A. Lorentz. Arriving more than half way through the book he states: "We could now enter upon a comparison of our dispersion formula with the measurements of the indices of refraction, but I shall omit this, because we must not attach too much importance to the particular form which we have found for the equations".

How bizarre! We will not know how well the final formula for the dispersion of light reproduces the measurements. This attitude is culminating in two review articles [6]. These are 120 pages about the different theoretical approaches which derive, with the help of quantum mechanics, the expression of the index of refraction. It is astonishing but there is not a single comparison between the theories and the indices of refraction data.

We know of two attempts about helium gas [7]. The authors tried an unnatural extension of the formula (3.2) where the ionisation of the atom enters the sum as a continuum and dominates that sum. Vinti supposes that the probability for the atom to be virtually ionised by a visible photon is 2.6 times larger than the probability to be excited. With Wheeler this ratio goes up to 3.

Indeed, all measurements of refraction indices are summarized through old empirical formulas:

$n_{\text{Cauchy}} = A + B\lambda^2 + C\lambda^4$, around 1840, $n_{\text{Sellmeier}} = \sqrt{1 + \frac{B_1 \lambda^2}{\lambda^2 - C_1} + \frac{B_2 \lambda^2}{\lambda^2 - C_2}}$, in 1871, and

$n_{\text{Hartmann}} = A + \frac{B}{\lambda - \lambda_0}$, by 1900. Most of the time, two terms are enough to reproduce the data.

Strangely enough the theoretical formulas (3.1), (3.2) or (3.3) are never used.

The physical origin for the reduced speed is thus very mathematical indeed, limited to dilute medium (but not too much), to plane waves established since the beginning of time and to molecules standing still. From this review we have to conclude that we lack of a proper interpretation of the slowing down of light in transparent mediums.

## 4. Our corpuscular approach

We propose a particle model for the index of refraction. It is based on a stop and go process. The photon has energy $E_\gamma$ and is propagating through molecules in their ground state.

Like in the Maxwell-quantum framework the molecules are described with quantum mechanics and they offer an infinite number of discrete excited energy states $W_i$ above their ground state $W_0$.

The photon, on the contrary, is very different from what it is in the Maxwell-quantum framework. We describe it as a particle with energy and a direction of propagation. It has no spatial extension or, at least, it is smaller than the atomic dimensions. There is no frequency, no wavelength and no electric field associated with our photon.

*4.1 Propagation through a gas*

The photon borrows an energy $\Delta E_i = E_i - E_\gamma$ in order to be absorbed by the molecule and to stop inside for a time: $\Delta t_i = \frac{\hbar}{\Delta E_i}$. After its stop, it gives back the energy $\Delta E_i$ and it resumes its trip with the speed c, the vacuum velocity. We will thereafter summarize these possibilities into a single average $<\Delta E>$ to be borrowed and a corresponding average stop time $<t_{\text{stop}}>$.

The stop time is given roughly by $c<t_{\text{stop}}> = \frac{2000 eV \dot{A}}{10 eV} = 200 \dot{A}$. It is therefore about 100 times as long as the time light takes to cross a few angstroms.

Let the mean free path between collisions be: $\Lambda$. The total average time to cross a distance $\Lambda$, is $<t_{\text{stop}}> + \Lambda/c$, and the average speed V is $\Lambda (<t_{\text{stop}}> + \frac{\Lambda}{c})^{-1}$.

Finally, the refractivity is

$$n - 1 = \frac{c <t_{\text{stop}}>}{\Lambda} \quad (4.1)$$

The photon has to propagate along a straight line and therefore the photon atom interaction is not an elastic process, otherwise the photon could be reemitted in other directions than the incident one. The process does not conserve energy anyway and it is reasonable to put back everything the way it was but for time. When the photon starts again it gets back the energy and the momentum that it had before the stop. The molecule stays still. Concerning the cross section for the photon to be absorbed momentarily by a molecule, we try the following: $\sigma \approx \alpha \pi R^2 = \alpha \sigma_\perp$. R is the radius of the molecule and $\alpha$ the fine structure constant characteristic of the low efficacy of light interactions. $\sigma_\perp$ is the geometric cross section of the molecule seen from the photon. The average volume that a molecule occupies is obtained from the mass density of the solid or the liquid, the mass of a mole and the Avogadro number.

The mean free path for a photon is then: $\Lambda = (\alpha \sigma_\perp N_{mol})^{-1}$. $N_{mol}$ is the number of molecules per unit volume. Finally our prediction for the refractivity is:

$$[n-1]_{corpuscular} = \alpha \sigma_\perp N_{mol} \hbar c \sum_i \frac{f_i}{E_i - E_\gamma} \quad (4.2)$$

We simplify this formula by defining an average molecular excitation energy $E_{AV}$ such that

$$\sum_i \frac{f_i}{E_i - E_\gamma} \approx \frac{1}{E_{AV} - E_\gamma}$$

Then we get:

$$[n-1]_{corpuscular} \approx N_{mol} \alpha \sigma_\perp \frac{\hbar c}{E_{AV} - E_\gamma} \quad (4.3)$$

*4.2 Propagation through a condensed medium*

The formula for the index of refraction in a condensed medium is going to be simplified with respect to what we have in a gas. In a crystal we have the molecules packed and touching each other. The numerical density of these molecules can be expressed in terms of the spacing between them.

If we have three axes Ox, Oy and Oz and the corresponding spacings: $\delta_x, \delta_y$ and $\delta_z$ then: $N_{mol} = (\delta_x \delta_y \delta_z)^{-1}$. If the photon is propagating along Ox we have:

$$\sigma_\perp = \delta_y \delta_z \implies \alpha \sigma_\perp N_{mol} = \frac{\alpha}{\delta_x}$$

$$n_x - 1 = \alpha \sigma_\perp N_{mol} \frac{\hbar c}{E_{AV} - E_\gamma} = \frac{\alpha}{\delta_x} \frac{\hbar c}{E_{AV} - E_\gamma} \quad (4.4)$$

This shows that having a cross section in our index formula leads to sensitivity to the spacing in the direction of propagation. When the molecule is very asymmetrical and fixed in position like in a crystal, there will be a varying index with the direction of the light in the crystal. In the Maxwell-quantum formula this does not happen because the number of valence electrons in a molecule does not depend upon the direction of propagation. This could be an important aspect of the physics behind the phenomenon of birefringence.

**5. Questions to the Maxwell-quantum model**

We list now five questions which are in fact criticisms to the Maxwell-quantum framework and which motivated the development of our corpuscular approach.

1- *What happens to light going through a gas with, for instance, less than one molecule per $\lambda^3$ ?*

This question concerns the limits in the use of the Maxwell theory of electromagnetism in macroscopic media. In our corpuscular approach the photon being a neutral object has no electric field and thus there is no problem since we do not use Maxwell theory of light.

2- *What do we do for condensed mediums?*

This is linked to the local field not being the incident field. The light is considered to be a wave and has therefore a spatial extent. This question is coming about because of this extended aspect of light. In our approach there is no problem of local fields versus incident fields since we do not have fields.

3- *If instead of an everlasting plane wave we envisage a more realistic burst of light it seems clear that a large part of the secondary waves will not have time to catch up with the incident light and will not therefore change the incident burst. Does the velocity appear to change as we move downstream?*

Again this is linked to the extended aspect of light in the classical theory and it disappears in our local photon theory.

*4- If instead of steady we consider moving molecules like in the Fizeau experiment with running water, then the molecules will reemit Doppler shifted waves with a frequency which will depend upon the angle of emission. Therefore the summation with the incident plane wave will be composed of different frequencies. Then the resulting apparent phase shift depends upon time! Is it reasonable to have an index of refraction which depends upon time?*

This is a mix between the extended aspect and the summing of the secondary wavelets to the main incident wave. We do not have the extended problem and our photons do not scatter, they keep going straight. Further it is well known that the Lorentz transformation applied to the photons in water gives back the results of Fizeau.

*5- Is it reasonable to imagine that the atomic electrons reemit light of the same frequency as the incoming wave when we know that atoms in a gas emit lines and not a continuum?*

This is a very serious concern for the Maxwell-quantum framework. The way we see it in our approach is that since a photon does not possess an electric field it does not set the atomic electrons into a periodic motion. In its way through matter, a photon is either absorbed momentarily by the molecule or it ignores the molecule. In any case the molecules do not emit photons on top of the propagating one.

**6. Comparisons between the Maxwell-quantum and the corpuscular expressions.**

We will compare the two formulas (3.2) and (4.3) in gases.
As we did for our corpuscular model we will approximate (3.2) by an average excitation energy $E_{AV}$.

$$[n-1]_{\text{Maxwell}} \approx N_{elec} 2\pi r_e \frac{(\hbar c)^2}{E_{AV}^2 - E_\gamma^2} \tag{6.1}$$

$$[n-1]_{\text{corpuscular}} \approx N_{mol} \alpha \sigma_\perp \frac{\hbar c}{E_{AV} - E_\gamma} \tag{6.2}$$

The number density $N_{elec}$ of dispersion electrons is not the number of electrons in a molecule (14 for $N_2$) times the number density of molecules $N_{mol}$. It is in fact the number of valence electrons: $n_{valence}$ times $N_{mol}$. $n_{valence}$ is the number of electrons missing to get to 8, a complete shell. For instance in $N_2$ this is $n_{valence} = 2 \times 3 = 6$. Since $N_{elec} = N_{mol} n_{valence}$ we have:

$$[n-1]_{\text{Maxwell}} \approx N_{mol} n_{valence} 2\pi r_e \frac{(\hbar c)^2}{E_{AV}^2 - E_\gamma^2} \tag{6.3}$$

The two formulas (6.2) and (6.3) differ both in the constants and in the energy dependences.

*6.1 Close to an atomic line*

We can imagine that if we get close to an atomic line, $E_0$, the data will show which form is right. In alkali vapours it is possible to get close to a line with visible light. For instance, within a few percent of the yellow doublet of Na, the data [8] are best fit with

$$[n-1]_{\text{data}} \propto \frac{\lambda_0}{\lambda - \lambda_0}$$

$$\lambda_0 = 0.589 \, 10^{-6} m, \quad \lambda_0 = \frac{2\pi \hbar c}{E_0}, \quad \lambda = \frac{2\pi \hbar c}{E_\gamma}$$

$$[n-1]_{\text{data}} \propto \frac{\lambda_0 E_\gamma}{2\pi \hbar c - \lambda_0 E_\gamma}$$

$$[n-1]_{\text{data}} \propto -1 + \frac{2\pi \hbar c}{2\pi \hbar c - \lambda_0 E_\gamma} = -1 + \frac{E_0}{E_0 - E_\gamma}$$

In [8] $E_\gamma$ is within a few percent of $E_0$. We can therefore drop (-1) and we get

$$[n-1]_{\text{data}} \propto \frac{E_0}{E_0 - E_\gamma}$$

This is exactly our corpuscular formula.

The Maxwell-quantum formula does well also since

$$[n-1]_{Maxwell} \propto \frac{1}{E_0^2 - E_\gamma^2} \approx \frac{1}{E_0 - E_\gamma} \frac{1}{2E_0}$$

Thus the two formulas show the same energy behaviour close to an atomic line.

*6.2 Far from an atomic line*
Let us compare the constants when $E_\gamma \ll E_{AV}$

$$[n-1]_{corpuscular} \approx N_{mol} \alpha \sigma_\perp \frac{\hbar c}{E_{AV} - E_\gamma}$$

$$[n-1]_{Maxwell} \approx N_{mol} n_{valence} 2\pi \frac{e^2}{m_e c^2} \frac{\hbar c}{E_{AV}} \frac{\hbar c}{E_{AV} - E_\gamma}$$

We treat the molecules exactly in the same way as in the Maxwell-quantum approach therefore $E_{AV}$ should be the same. If we want the refractivity to be the same we have to have:

$$N_{mol} \alpha \sigma_\perp \frac{\hbar c}{E_{AV} - E_\gamma} \approx N_{mol} n_{valence} 2\pi\, r_e \frac{e^2}{m_e c^2} \frac{\hbar c}{E_{AV}} \frac{\hbar c}{E_{AV} - E_\gamma}$$

$$\sigma_\perp \approx n_{valence} 2\pi \frac{\hbar c}{m_e c^2} \frac{\hbar c}{E_{AV}}$$

$n_{valence}$ is a few units, and $E_{AV}$ is between 10 and 20 eV. These numbers imply that the radius of the molecule is a few Angstrom, which is correct. This shows that the two formulas are not orders of magnitude away from each other. However, in the next paragraph we show that, in order to reproduce the data, the Maxwell-quantum $E_{AV}$ has to be larger than the $E_{AV}$ of the corpuscular approach. Further, in the condensed phase, the Maxwell-quantum $E_{AV}$ is much larger than the typical excitation energies of the medium.

## 7. Comparisons to experimental data far from resonances

One problem with the light wave or Maxwell-quantum framework is the fact that there is no unique way to define the speed of a wave packet. At least five velocities can be defined for a wave packet: phase, group, front, signal and energy velocities [9].

It has been shown experimentally [10] that a single photon with energy far from atomic resonances, travels at the group velocity. In our corpuscular approach there is a statistical distribution of the number of stops but there is a single, definite, average velocity and it is, obviously, the photon velocity. Therefore our corpuscular formula (4.3) is to be compared with the group index data. The theoretical Maxwell-quantum index (3.1, 3.2 and 3.3) is obtained through the phase shift of the incident plane wave thus it corresponds to the phase index: $n_p$. We determine the group index: $n_g$, both for the Maxwell-quantum formula and for the experimental data, through the following relationship:

$$n_g = \frac{n_p}{1 - \frac{E_\gamma}{n_p} \frac{dn_p}{dE_\gamma}} \quad (7.1)$$

We will consider the experimental data about Air and fused silica group indices.

*7.1 Air*
For many purposes like astronomy, GPS communications and light detection and ranging (LIDAR), the knowledge of the refractive index of air is very important. The number of measurements is large and this is the reason why we choose air as an example. The group index data are taken from [11] with the following conditions: dry air, 15°C, 450ppm $CO_2$, 1013 hPa => $N_{mol} = 2.55\, 10^{25}$ molecules/m$^3$.
In our corpuscular model we need to estimate the geometric cross section $\sigma_\perp$ of the air molecules.
$N_2$ => $6\, 10^{23}$ molecules weigh 28g and the liquid has a density of 808.6 kg/m$^3$. A single molecule occupies a volume: $28/(6\, 10^{23}\, 8.09\, 10^5) = 57.7\, 10^{-30}$ m$^3$
The transverse area is thus estimated as $\pi/4\, (V_{N2})^{2/3} = 11.7\, 10^{-20}$ m$^2$.
$O_2$ => $6\, 10^{23}$ molecules weigh 32g and the liquid has a density of 1141 kg/m$^3$. A single molecule occupies a volume: $32/(6\, 10^{23}\, 11.41\, 10^5) = 46.7\, 10^{-30}$ m$^3$

The transverse area is thus estimated as $\pi/4 \ (V_{O2})^{2/3} = 10.2 \ 10^{-20}$ m$^2$.
We calculate then the weighted average transverse area for Air : $<\sigma_\perp> = 11.4 \ 10^{-20}$ m$^2$.

$$[n_g - 1]_{corpuscular} = \frac{2.55}{137} 10^{25} 11.4 \ 10^{-20} \frac{197 \ 10^{-9} eVm}{E_{AV} - E_\gamma} = 10^{-3} \frac{4.18}{E_{AV} - E_\gamma}$$

In the Maxwell-quantum model we need the number density of the valence electrons. For $N_2$, $n_{valence} = 2 \times 3$ and for $O_2$, $n_{valence} = 2 \times 2$. For Air this will average to $0.8*6+0.2*4 = 5.6$ valence electrons per air molecule. Then under our Air conditions we get: $N_{elec} = 14.28 \ 10^{25}$ valence electrons/m$^3$. As for the corpuscular formula we try a one term approach with an average energy. The phase refractivity is then approximately

$$[n_P - 1]_{Maxwell} = \frac{98.5 \ 10^{-3}}{E_{AV}^2 - E_\gamma^2}$$

The energies $E_{AV}$ and $E_\gamma$ are in eV.
Then we make use of (7.1) to get the group index of the Maxwell-quantum approach.

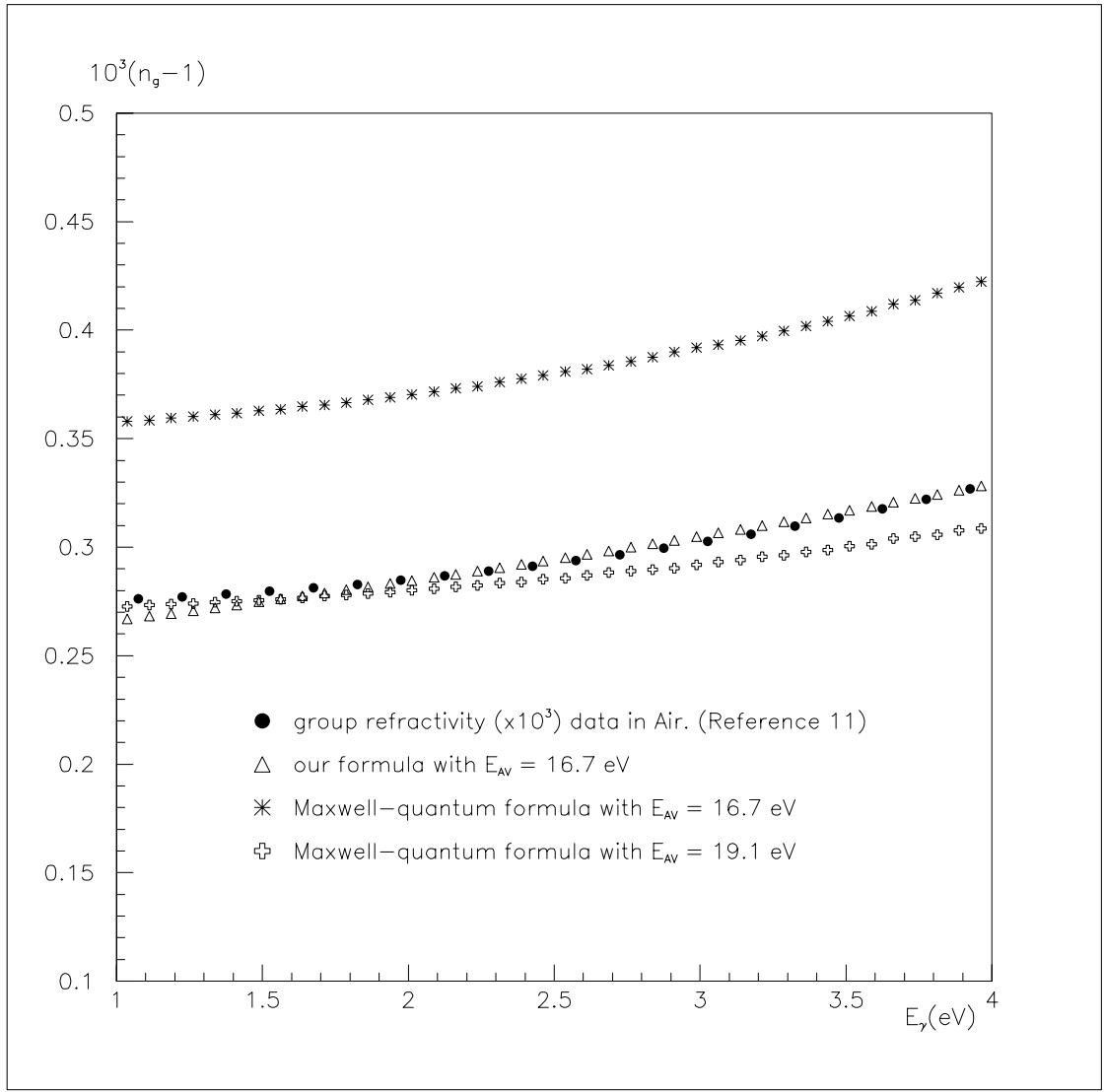

Figure 1. $(n_g-1)*10^3$ for Air, as a function of the photon energy in eV.

The figure 1, shows the Maxwell-quantum prediction with $E_{AV} = 16.7$ eV. It is clear that it needs to be reduced. When $E_{AV} = 19$ eV, the group index values are fine but the Maxwell-quantum curve is then too flat. We conclude that the two predictions for Air are not very different except that the Maxwell-quantum formula demands the average excitation energy to be larger than in the corpuscular formula.

*7.2 Fused silica*

The data for synthetic fused silica, $SiO_2$, were found in [12]. The weigh of $6\ 10^{23}$ molecules is 60.1 g and the density at 25°C is 2200 kg/m$^3$. Thus a molecule occupies a volume: $60.1/(6\ 10^{23}\ 22\ 10^5) = 45.4\ 10^{-30}$ m$^3$. In other words, there are $2.2\ 10^{28}$ molecules per m$^3$.

The spacing is: $\delta_{SiO2} = (45.4\ 10^{-30})^{1/3}\ m = 3.57\ 10^{-10}\ m$, and the corpuscular prediction is

$$[n_{gSiO2} - 1]_{corpuscular} = \frac{\alpha}{\delta_{SiO2}} \frac{\hbar c}{E_{AV} - E_\gamma} = \frac{4.08\ eV}{E_{AV} - E_\gamma}$$

In the Maxwell-quantum framework we use the Lorenz-Lorentz formula (3.3).
The number of valence electrons for $SiO_2$ is $4+2\times2 = 8$
The number density is: $2.2\ 10^{28} * 8 = 17.6\ 10^{28}$ valence electrons/m$^3$.

$$\frac{n_P^2 - 1}{n_P^2 + 2} = \frac{4\pi}{3} 2.83\ 10^{-15} 17.6\ 10^{28} 3.88\ 10^{-14} \frac{1}{E_{AV}^2 - E_\gamma^2} = \frac{81}{E_{AV}^2 - E_\gamma^2} = A$$

$$[n_p - 1]_{LL} = \sqrt{\frac{1+2A}{1-A}} - 1$$

Lorentz explains in his book that formula (3.3) is not always working fine and that the old Laplace formula (3.2) is to be considered too because experimental data sometimes stand in between. So we will also draw the refractivity obtained with Laplace formula:

$$[n_p - 1]_{Laplace} = \sqrt{1+3A} - 1$$

Then we calculate $n_g$ from $n_p$ with formula (7.1).

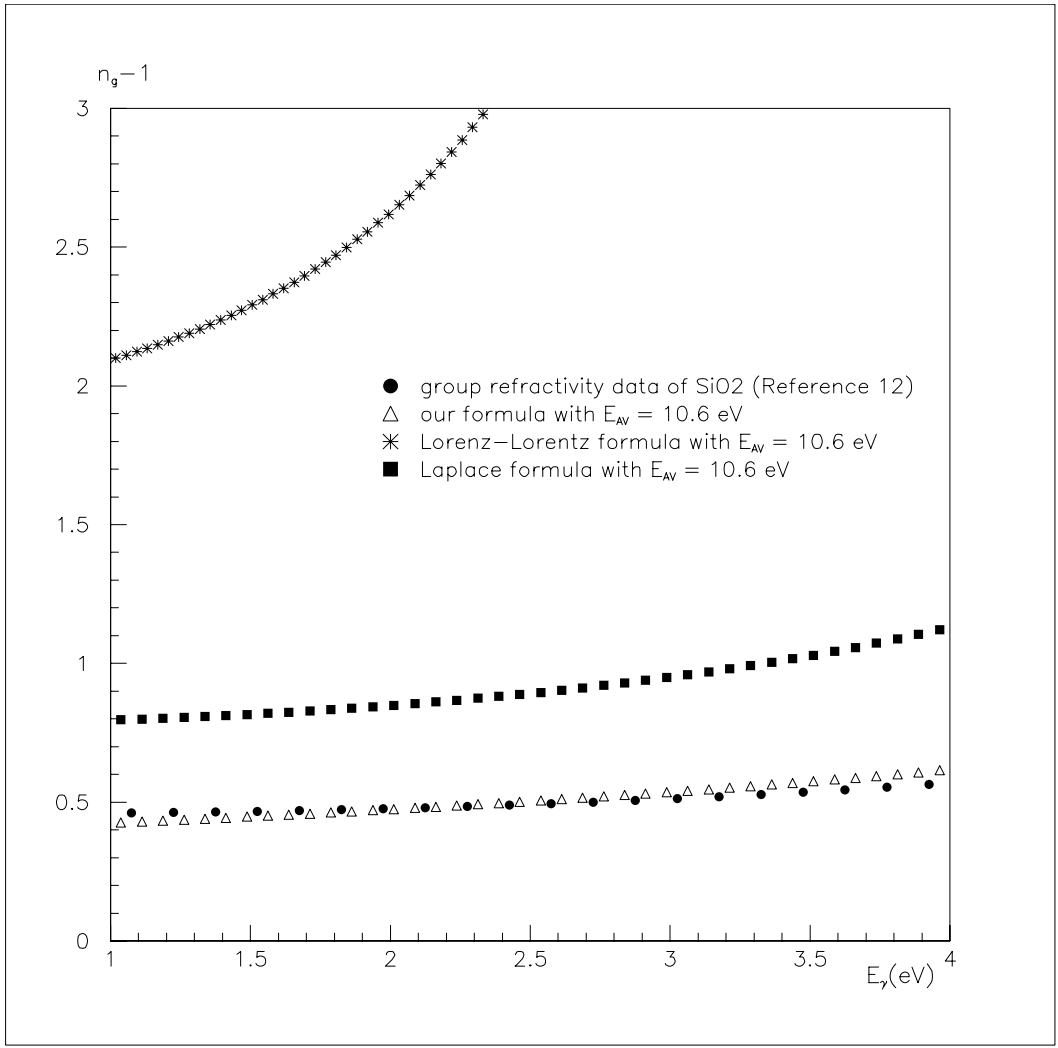

Figure 2. The fused silica group refractivity: $n_g$-1, as a function of the photon energy in eV.

The figure 2, shows the data together with our simplified formula with $E_{AV}$ = 10.6 eV. The UV absorption band in $SiO_2$ begins at 0.15 $\mu m$ (8.2 eV) so an average excitation energy of about 10 eV is reasonable. Similarly to what we did with the Air data we show the Maxwell-quantum prediction with the same average excitation energy of 10.6 eV.

Clearly, in order to get a reasonable value for the silica refractivity, the Maxwell-quantum predictions need higher average excitation energy than for Air data. We have to plug a value of 17.3 eV for the average excitation energy in the Lorenz-Lorentz formula and 14.8 eV in the Laplace one. We understand now why Vinti and Wheeler had to weigh heavily the ionization energies in order to lower sufficiently the index of refraction of helium.

We conclude by saying that the situation is bad for the Maxwell-quantum predictions in a condensed medium while it looks good for our corpuscular approach.

**8. Distribution of the traversal times of a photon through a crystal**

In our corpuscular model, the time, for a photon, to cross a given thickness of transparent matter is not a constant. The time dispersion is given by the statistical fluctuations of the number of photon-molecule encounters which lead to a stop.

In the Maxwell-quantum model the inverse of the bandwidth gives the coherence time which, multiplied by c, is considered as the length of the photon. The detection of a photon will occur along that length. This leads to a distribution of traversal times although it is independent of the matter being traversed.

Independently of the model choosen, the dispersion of the index of refraction, roughly proportionnal to the dispersion in photon energies, does produce a scatter of the traversal times.

It turns out that we may already have a hint about these phenomenons in the results of a remarkable experiment [13]. A KDP crystal pumped with a 351nm laser was producing, through the parametric down-conversion effect, a signal and an idler photon. They were selected to have the same energy ( $\lambda = 700 nm \Rightarrow E_\gamma = E_0 \approx 1.8\,eV$ ) and sent on each side of a beam splitter. The time coincidences between the two outputs of the beam splitter were then registered as a function of the difference in path length. The boson character of the photons implies that when they arrive at the same time at the beam splitter and with the same energy, they leave the beam splitter in the same state. This means that they both end up in the same detector and the coincidence rate drops to zero. The width of the coincidence dip is interpreted as the time dispersion between the two photons in the pair.

We examine now the possible origins for that time dispersion.

*8.1 effect of the bandwidth*

Interference filters of about 7nm were inserted both in the signal and in the idler path. As $\frac{\Delta\lambda}{\lambda} = \frac{7}{700} = \frac{\Delta f}{f}$ we conclude that: $\Delta f = \frac{f}{100} = 4\,10^{12}\,Hz$. This, if light is a wave, implies a time dispersion such that:

$$FWHM\,(t_{traversal}, 1\gamma) > \frac{1}{\Delta f} = 250\,fs$$

The experimental set up had also two pinholes that could have further restricted the energy dispersion of the photons and that explain why the FWHM could be larger. The time difference between the signal and the idler photon will be given by the folding of the two distributions. This leads to adding the variances and thus:

$$FWHM\,(t_{traversal}, 2\gamma) > \frac{\sqrt{2}}{\Delta f} = 350\,fs$$

*8.2 effect of the dispersion*

When the energy of the photon is such that: $E_0 - \varepsilon < E_\gamma < E_0 + \varepsilon$, the energy dispersion will change the traversal times as:

$$FWHM(t_{traversal}) = 2\varepsilon \frac{d}{dE\gamma}\left(\frac{L}{V(E_\gamma)}\right) = 2\varepsilon \frac{L}{c}\frac{dn(E_\gamma)}{dE\gamma}$$

The energy bandwidth is given by: $\frac{\Delta E_\gamma}{E_\gamma} = \frac{\Delta\lambda}{\lambda} = \frac{7}{700}$, $\Delta E_\gamma < 0.018\,eV$ and $\varepsilon < 0.009\,eV$

Experimentally [14]: $n_{KDP}(E_\gamma \approx 2eV) = 1.5$ and $\frac{dn_{KDP}(E_\gamma \approx 2eV)}{dE\gamma} = 1.3\,10^{-2}(eV)^{-1}$

$$FWHM(t_{traversal}) = 86\,10^{-12}\,L\varepsilon$$

This is linear with the thickness of the crystal and with the energy bandwidth of the photons.

$$FWHM(t_{traversal}) < 730\,L\,fs$$

The length of the KDP crystal in this experiment is 8cm and the pair production is a rare process, thus the average thickness is <L> = 4cm. We end up with

$$FWHM(t_{traversal}) < 30\,fs$$

*8.3 effect of the statistics of the number of stops*

In our corpuscular model, the average number of stops follows a Poisson law.

We will assume that $\delta_{KDP} \approx 3.6\,10^{-10}\,m$ and $E_{AV} \approx 12\,eV$

The average number of stops is:

$$N_{stop} = \frac{L}{\Lambda_{KDP}} = \alpha\frac{L}{\delta_{KDP}} = 2\,10^7\,L$$

And the average stop-time:

$$<t_{stop}> = \frac{\hbar}{E_{AV} - E_\gamma} \approx 60\,10^{-18}\,s$$

$$FWHM(t_{traversal},1\gamma) = 2.35<t_{stop}>\sqrt{N_{stop}} = 2.35\,60\,10^{-15}\,4.5\sqrt{L} = 630\sqrt{L}\,fs$$

This is proportional to the square root of the thickness.
Again with an average <L> = 4cm we have

$$FWHM(t_{traversal},1\gamma) \approx 120\,fs$$

We have to fold the two Gaussian distributions since we have two photons and this is

$$FWHM(t_{traversal},2\gamma) \approx \sqrt{2}\,120\,fs = 170\,fs$$

*8.4 Comparison with the experimental result* [13]

The observed FWHM is 56fs but it has to be doubled because when the beam splitter moves so as to reduce one path length it increases at the same time, the other path. Further if the beam splitter is attacked with a 45° incident angle there is a further multiplicative factor 1/cos(45°). The real time FWHM turns out to be 56*2*1.414 = 158 fs. The effect of the bandwidth is predicted to be larger than 350 fs. The effect of the dispersion is less than 30 fs. The only effect which corresponds to the measurements is our statistical stop time prediction. We conclude that light does not seem to be a wave and that it may very well be that our prediction of the statistical distribution of traversal times has been already seen in 1987!

**9. Understanding the optical Kerr effect**

An intense beam of light produces, in a medium, an increase of its refractive index [15], [16], [17], [18], [19], [20]. This phenomenon has been observed in solids, liquids and gases and is known by the names of AC Kerr, optical Kerr or self focusing effect. This is usually expressed empirically with the formula: $n = n_0 + n_2 I$. $n_0$ is the usual index of refraction at low light intensity I. The constant $n_2$ is comprised between $10^{-20}$ m$^2$/W for dense media and $10^{-23}$ m$^2$/W for gases.

*9.1 The physics of the Kerr effect in our corpuscular approach*

A photon crossing a molecule has a probability $\alpha$ to stop. If the photon stops and if we have, at the same time, $1/\alpha$ photons on top of the first, we will have on average two photons stopping in the molecule. The stop time for a photon is proportional to $1/\Delta E$ where $\Delta E$ is the energy borrowed to get from $E_\gamma$ to $E_1$. These two photons need only borrow $\dfrac{E_1 - 2E_\gamma}{2}$ each. When $E_1 \gg E_\gamma$, this is roughly half the value for a single photon. Therefore they will stop for a time twice as long and since the refractivity: ($n_0$ -1), is proportional to the stop time, we expect a doubling of the refractivity.

Thus when the density of photons is $1/\alpha$ in a molecular volume, the refractivity is roughly multiplied by a factor two. Let us consider visible photons of energy 1 eV propagating through water, and estimate the energy flux necessary to get such a density. A water molecule occupies approximately, a volume of 30 $10^{-30}$ m$^3$.

$$\rho_{photons} = \frac{1/\alpha}{V_{molecule}} = \frac{137}{30\,10^{-30}} = 4.6\,10^{30}\ photons/m^3$$

$$\Phi_{photons} = \rho_{photons}\frac{c}{n} = 4.6\,10^{30}\,\frac{3\,10^8}{1.33} \approx 10^{39}\ photons/m^2/s$$

This can be translated into an energy flux

$$I\ (W/m^2) = \Phi_{photon} E_\gamma = 10^{39}\,1.6\,10^{-19} = 1.6\,10^{20}\ W/m^2$$

So we understand three things:
1- the index of refraction increases when the flux of photons increases.

$$n = n_0 + n_2 I$$

2- $n_2 \approx 10^{-20}$ m$^2$/W in water.
3- the refractivity: n – 1, is proportional to the number density of incoming photons and is also roughly proportional to the density of the medium. We understand therefore why $n_2$ is about 1000 times smaller for gases than for condensed mediums.

*9.2 The prediction for $n_2$ and the comparison to the data*

Consider an energy flux I (W/m$^2$) of photons of energy: $E_\gamma$, falling on a slab of matter of thickness L. The number density of photons in the slab is $\rho_{photons}$, and we have: $\Lambda = (\alpha \sigma_\perp N_{mol})^{-1}$ (4.1).

When $\rho_{photons} V_{molecule} \ll 1/\alpha$, we consider two possibilities only:
- a single photon stops in a molecule and stays for a time $\Delta t_1$
- two photons stop in a molecule and stay for a time $\Delta t_2$.

If $L \ll \Lambda$ the average number of photons which stop per m$^2$ and per second is:

$$N_1 \approx \frac{L}{\Lambda}\frac{I}{E_\gamma} = \alpha \sigma_\perp N_{mol} L \frac{I}{E_\gamma}$$

they will stop for a time

$$\Delta t_1 = \frac{\hbar}{E_{AV} - E_\gamma}$$

A fraction of these: $N_2 = N_1 \alpha \rho_{photons} V_{molecule}$, will be absorbed together with a companion and this corresponds to a stop time of

$$\Delta t_2 = \frac{\hbar}{(E_{AV} - 2E_\gamma)/2} = \frac{2\hbar}{E_{AV} - 2E_\gamma}$$

The stop time $\Delta t_2$ affects two photons.
This leads to an average stop time which is given by:

$$<\Delta t> = \frac{N_1 - N_2}{I/E_\gamma}\Delta t_1 + \frac{N_2}{I/E_\gamma}2\Delta t_2 = \frac{E_\gamma}{I}\left[N_1 \Delta t_1 + N_2(2\Delta t_2 - \Delta t_1)\right]$$

$$<\Delta t> = \frac{L}{\Lambda}\frac{\hbar}{E_{AV} - E_\gamma}\left[1 + \alpha \rho_{photon} V_{molecule}\frac{3E_{AV} - 2E_\gamma}{E_{AV} - 2E_\gamma}\right]$$

The total time spent crossing the distance L is: $L/c + <\Delta t>$.

The average speed is thus $<V> = \frac{L}{L/c + <\Delta t>}$ and the index n, is c/<V>:

$$n - 1 = \frac{c <\Delta t>}{L} = \frac{1}{\Lambda} \frac{\hbar c}{E_{AV} - E_\gamma} \left[ 1 + \alpha \rho_{photon} V_{molecule} \frac{3E_{AV} - 2E_\gamma}{E_{AV} - 2E_\gamma} \right]$$

$$\rho_{photon} = \frac{n_0}{c} \frac{I}{E_\gamma}$$

We recognise our expression for the usual refractivity (see §4):

$$n_0 - 1 = \frac{1}{\Lambda} \frac{\hbar c}{E_{AV} - E_\gamma}$$

$$n = n_0 + (n_0 - 1) \frac{n_0}{c} \alpha V_{molecule} \frac{3E_{AV} - 2E_\gamma}{E_{AV} - 2E_\gamma} \frac{1}{E_\gamma} I = n_0 + n_2 I$$

$$n_2 = (n_0 - 1) \frac{n_0}{c} \alpha V_{molecule} \frac{3E_{AV} - 2E_\gamma}{E_{AV} - 2E_\gamma} \frac{1}{E_\gamma}$$

We can now calculate $n_2$ for different materials and compare with the experimental data.

In water: $V_{molecule} = 30 \, 10^{-30} m^3$, $n_0 = 1.33$, $E_{AV} \approx 8$ eV

$$n_{2\,Water} = 3.3 \, 10^{-1} \frac{1.33}{3 \, 10^8} \frac{1}{137} 30 \, 10^{-30} \frac{3E_{AV} - 2E_\gamma}{E_{AV} - 2E_\gamma} \frac{1}{1.6 \, 10^{-19}} \approx 0.75 \, 10^{-20} \, m^2/W$$

This is in agreement with the experimental measurements [15], [16], [17], [18], which are between 0.5 and 5 $10^{-20}$ m$^2$/W

In Air: $V_{molecule} = 55 \, 10^{-30} m^3$, $n_0 - 1 = 3 \, 10^{-4}$, $E_{AV} = 16.7$ eV

$$n_{2\,Air} = 3 \, 10^{-4} \frac{1}{3 \, 10^8} \frac{1}{137} 55 \, 10^{-30} \frac{3E_{AV} - 2E_\gamma}{E_{AV} - 2E_\gamma} \frac{1}{E_\gamma} \approx 0.8 \, 10^{-23} \, m^2/W$$

It is a bit smaller than the experimental findings [19], [20], which scatter, with large errors, between 4 $10^{-23}$ and 12 $10^{-23}$ m$^2$/W. We can understand that the effective area of the laser beam is not an easy quantity to measure.

Finally the optical Kerr effect is not, in our point of view, connected to a non linearity of the interaction light-medium.

## 10. Conclusions

We have shown that the Maxwell-quantum framework does not satisfactorily explain the very basic phenomenon of the slowing down of light in a transparent medium. We showed that a corpuscular approach explains naturally this fact and that it leads to easy and satisfactory predictions of the refractive indices.

In our approach there is no need to associate some frequency or wave length to the photon. We do not need either an electric field of the photon and this is fortunate because the photon being a neutral elementary object cannot be at the origin of an electric field. Indeed the only variables we need, in this approach, to describe the photon, are its energy, its direction of propagation and its dimension way smaller than the atomic sizes.

Another difference with the usual electro-magnetic wave or Maxwell-quantum approach is the importance of the cross section of molecules to visible photons. The geometric cross section of a very asymmetrical molecule depends upon its orientation with respect to the direction of propagation of the photon. This allows our index to depend upon that direction of propagation. This is not possible in the Maxwell-quantum framework which is sensitive to the volume density of electrons independent of the direction of travel.

In our model the measured speed of the photon which traverses a transparent material is a statistical process and we predict a Gaussian distribution of the crossing times. We believe that this prediction was present in the results of an experiment on photons from parametric down conversion.

At last our corpuscular description leads to new viewpoints and for instance we understand and predict the magnitudes of the optical Kerr effect both in gases and in condensed matter.

**Acknowledgements**

I wish to thank my colleagues: Barrand G, Haïssinski J and Zomer F, for numerous and helpful discussions.

**References**
[1] Foucault L 1850 *CR Acad. Sci. Paris* **30** 551
[2] Feynman R P, Leighton R B, and Sands M L 1965 *The Feynman Lectures on Physics* vol 1 (Addison-Wesley, New York)
[3] van de Hulst H C 1981 *Light Scattering by Small Particles* (Dover publications, New York) sections 4.3 and 4.5
[4] Lorentz H A 1952 *The Theory of Electrons*, second edition, (Dover publications, New York) Item 132 p 152
   Sommerfeld A 1954 *Optics*, Lectures on Theoretical Physics, vol 4 (Academic Press Inc) pp 91 and 92
   Rosenfeld L 1951 *Theory of Electrons* (North Holland Publishing Company, Amsterdam)
   Jackson J D 1975, *Classical Electrodynamics*, second edition, (John Wiley & sons)
   Panofsky W K H and Phillips M 1955 *Classical Electricity and Magnetism* (Addison-Wesley)
   Born M and Wolf E 1999 *Principles of Optics*, seventh edition, (Cambridge University Press)
[5] Koch J 1909 *Nova Acta Upsal.* 2
[6] Breit G 1932 Quantum theory of dispersion *Reviews of Modern Physics* **4** 504
   Breit G 1933 Quantum theory of dispersion *Reviews of Modern Physics* **5** 91
[7] Vinti J P 1932 *Phys. Rev.* **42** 632
   Wheeler J A 1933 *Phys. Rev.* **43** 258
[8] Korff S A 1929 *Phys. Rev.* **34** 457
[9] Brillouin L 1960 *Wave Propagation and Group Velocity* (Academic Press Inc. New York)
[10] Steinberg A M, Kwiat P G, and Chiao R Y 1992 *Phys. Rev. Let.* **68** 2421
[11] Ciddor P E and Hill R J 1999 *Applied Optics* **38** 1663
[12] Malitson I H 1965 *J. Opt. Soc. Am.* **55** 1205
[13] Hong C K, Ou Z Y and Mandel L 1987 *Phys. Rev. Let.* **59** 2044
[14] Zernike F 1964 *J. Opt. Soc. Am.* **54** 1215
[15] Maker P D, Terhune R W and Savage C M 1964 *Phys. Rev. Let.* **12** 507
[16] Ho P P and Alfano R R 1979 *Phys. Rev.* **A 20** 2170
[17] Gong Q H, Li J L, Zhang T Q and Yang H 1998 *Chin. Phys. Let.* **15** 30
[18] Kolesic M, Wright E M and Moloney J V 2004 *Phys. Rev. Let.* **92** 253901
[19] Hellwarth R W, Pennington D M and Henesian M A 1990 *Phys. Rev.* **A 41** 2766
[20] Nibbering E T J, Grillon G, Franco M A, Prade B S and Mysyrowicz A 1997 *J. Opt. Soc. Am.* **B 14** 650